# A Complex Constrained Total Variation Image Denoising Algorithm with Application to Phase Retrieval

Yunhui Gao and Liangcai Cao

*Abstract*—This paper considers the constrained total variation (TV) denoising problem for complex-valued images. We extend the definition of TV seminorms for real-valued images to dealing with complex-valued ones. In particular, we introduce two types of complex TV in both isotropic and anisotropic forms. To solve the constrained denoising problem, we adopt a dual approach and derive an accelerated gradient projection algorithm. We further generalize the proposed denoising algorithm as a key building block of the proximal gradient scheme to solve a vast class of complex constrained optimization problems with TV regularizers. As an example, we apply the proposed algorithmic framework to phase retrieval. We combine the complex TV regularizer with the conventional projection-based method within the constraint complex TV model. Initial results from both simulated and optical experiments demonstrate the validity of the constrained TV model in extracting sparsity priors within complex-valued images, while also utilizing physically tractable constraints that help speed up convergence.

*Index Terms*—Convex Optimization, Proximal Algorithms, Total Variation, Image Denoising, Phase Retrieval, Complex-Valued Images

## I. INTRODUCTION

TOTAL variation (TV) has remained a popular model for image processing since its introduction by Rudin *et al.* [1]. It has been found particularly suitable for characterizing natural images because of the edge-preserving property of the TV regularization [2]–[4]. It has been widely studied for image restoration tasks [5]–[7], together with many variants, such as color TV [8], bilateral TV [9], weighted TV [10], and adaptive TV [11], etc. Other researchers focus on the algorithmic side, providing numerical solvers for the corresponding optimization problems. For example, see [12]–[16] and references therein.

Despite its success in dealing with real-valued images, TV models in the complex domain remain largely unexplored. Complex-valued signals arise in many areas of imaging science, such as optics [17], acoustics [18], radar [19], X-ray imaging [20], and magnetic resonance imaging [21], etc. This is largely due to the introduction of complex notation as a convenient mathematical tool for wavefield representation. For these imaging applications, recovering the complex-valued images usually requires solving an ill-posed optimization problem. Numerical reconstruction relies heavily on the regularization scheme that exploits prior knowledge of the unknown signals, indicating potential applications of the TV models.

In this paper, we extend the TV regularization to the complex domain. Specifically, we aim to solve the constrained denoising problem:

$$\min_{\mathbf{x}\in C} \; \frac{1}{2}\|\mathbf{x}-\mathbf{b}\|_F^2 + \lambda \|\mathbf{x}\|_{\text{TV}}, \tag{1}$$

where $\mathbf{x}$ represents the complex-valued image to be recovered, $\mathbf{b}$ is a known and possibly noisy observation, $\|\cdot\|_{\text{TV}}$ is a discrete TV seminorm, $\lambda > 0$ is a regularization parameter, and $C$ is a closed convex set. The reason why this problem is important is that, with the help of the proximal gradient method [22], solving problem (1) usually suffices to find solutions to a wider range of constrained minimization problems of the form:

$$\min_{\mathbf{x}\in C} \; F(\mathbf{x}) + \tau \|\mathbf{x}\|_{\text{TV}}, \tag{2}$$

where $F$ is a smooth function, and $\tau > 0$ is another parameter. In fact, many image restoration problems can be expressed in the form of (2).

One line of research that is of particular interest to our work is the dual approach to problem (1), which was first introduced by Chambolle to the unconstrained real-valued case [23]. Later in [24], Beck *et al.* proposed an accelerated gradient projection algorithm for the constrained case. The algorithm shares great simplicity together with a fast rate of convergence. Following their footsteps, we generalize the method to complex-valued images and derive the corresponding algorithm.

To demonstrate the effectiveness of our model and algorithm, we applied them to phase retrieval, a well-known optimization problem in imaging science [25]. In optical settings specifically, phase retrieval refers to the problem of recovering the complex-valued object from the magnitude of its diffraction pattern. The ill-posed nature of the problem renders algorithmic recovery

This paragraph of the first footnote will contain the date on which you submitted your paper for review. It will also contain support information, including sponsor and financial support acknowledgment. This work was supported in part by the National Natural Science Foundation of China under Grant 61775117 and Grant 61827825.

Y. Gao and L. Cao are with the State Key Laboratory of Precision Measurement Technology and Instruments, Department of Precision Instrument, Tsinghua University, Beijing 100084, China. (e-mail: gyh21@mails.tsinghua.edu.cn; clc@tsinghua.edu.cn).

rather challenging [26]. In light of this, we explored sparsity in the complex transmittance of real-world objects and tackled the inherent ill-posedness of the problem via TV-based regularizers.

The remainder of this paper is organized as follows. Section II reviews the real TV seminorms and presents the definition of complex TVs. Section III introduces the proximal gradient method and its accelerated variant, which provide a general algorithmic framework for solving nonsmooth optimization problems including (2). Section IV derives the complex constrained total variation denoising algorithm, which serves as a key building block for the proximal gradient method. In Section V, the proposed algorithm is applied to phase retrieval as an example. Numerical results are presented in Section VI, which is followed by some concluding remarks in Section VII.

*Notations:* Throughout the paper, we use boldface lowercase letters for vectors and matrices. The entries of a vector or a matrix are indexed by subscript letters. Boldface numbers **0** and **1** stand for the zero vector / matrix and all-ones vector / matrix respectively. Scalars, scalar-valued functions and sets are all denoted by italic letters, and should be distinguished according to the context. Cursive letters represent abstract operators. $\|\cdot\|_F$ and $\|\cdot\|_{\text{TV}}$ stand for the Frobenius norm and a TV seminorm, respectively. $\langle\cdot,\cdot\rangle$ denotes the inner product. $|\cdot|$, $(\cdot)^2$, $\sqrt{\cdot}$, and $\cdot/\cdot$ are element-wise operators, and $\odot$ stands for the element-wise (Hadamard) multiplication operator for vectors or matrices. When applied to vectors or matrices, the equality and inequality signs should also be interpreted as element-wise symbols. $\text{diag}(\cdot)$ operator puts the entries of a vector into the diagonal of a matrix.

## II. COMPLEX TOTAL VARIATION MODELS

Before introducing complex TVs, we first briefly review the definition of the TV seminorms for real-valued images. For an image $\mathbf{x} \in \mathbb{R}^{m \times n}$, there are two popular variants of TV [3], namely the isotropic TV defined by

$$\|\mathbf{x}\|_{\text{TVi}} = \sum_{j=1}^{m-1}\sum_{k=1}^{n-1} \sqrt{(x_{j,k} - x_{j+1,k})^2 + (x_{j,k} - x_{j,k+1})^2} \\ + \sum_{j=1}^{m-1} |x_{j,n} - x_{j+1,n}| + \sum_{k=1}^{n-1} |x_{m,k} - x_{m,k+1}|, \quad (3)$$

and the anisotropic TV defined by

$$\|\mathbf{x}\|_{\text{TVa}} = \sum_{j=1}^{m-1}\sum_{k=1}^{n-1} \{|x_{j,k} - x_{j+1,k}| + |x_{j,k} - x_{j,k+1}|\} \\ + \sum_{j=1}^{m-1} |x_{j,n} - x_{j+1,n}| + \sum_{k=1}^{n-1} |x_{m,k} - x_{m,k+1}|. \quad (4)$$

Both types of TV require computing the horizontal and vertical finite differences of the image but are different in their ways of summation. The anisotropic TV is not rotationally invariant as opposed to the isotropic one. As a result, the two TVs may favor different solutions.

For a complex-valued image $\mathbf{x} \in \mathbb{C}^{m \times n}$, let $\mathbf{u} \in \mathbb{R}^{m \times n}$ and $\mathbf{v} \in \mathbb{R}^{m \times n}$ denote the real and imaginary parts of $\mathbf{x}$, respectively. That is, we have $\mathbf{u} = \mathcal{R}(\mathbf{x})$, $\mathbf{v} = \mathcal{I}(\mathbf{x})$, and $\mathbf{x} = \mathbf{u} + i\mathbf{v}$, where $\mathcal{R}(\cdot)$ and $\mathcal{I}(\cdot)$ are element-wise operators that extract the real part and the imaginary part, respectively. The vertical and horizontal finite differences of $\mathbf{u}$ are denoted by $\mathbf{u}^{(1)} \in \mathbb{R}^{(m-1) \times n}$ and $\mathbf{u}^{(2)} \in \mathbb{R}^{m \times (n-1)}$, which are defined as

$$\begin{aligned} u_{j,k}^{(1)} &= u_{j,k} - u_{j+1,k}, \quad j=1,\dots,m-1, k=1,\dots,n \\ u_{j,k}^{(2)} &= u_{j,k} - u_{j,k+1}, \quad j=1,\dots,m, k=1,\dots,n-1. \end{aligned} \quad (5)$$

Similarly, $\mathbf{v}^{(1)} \in \mathbb{R}^{(m-1) \times n}$ and $\mathbf{v}^{(2)} \in \mathbb{R}^{m \times (n-1)}$ are the vertical and horizontal finite differences of $\mathbf{v}$, which are defined as

$$\begin{aligned} v_{j,k}^{(1)} &= v_{j,k} - v_{j+1,k}, \quad j=1,\dots,m-1, k=1,\dots,n \\ v_{j,k}^{(2)} &= v_{j,k} - v_{j,k+1}, \quad j=1,\dots,m, k=1,\dots,n-1. \end{aligned} \quad (6)$$

In this work, we consider two types of complex TV. The first type, which we refer to as type-I, is a direct extension from the real-valued case. The type-I isotropic TV is defined as

$$\begin{aligned} \|\mathbf{x}\|_{\text{TV1i}} &= \sum_{j=1}^{m-1}\sum_{k=1}^{n-1} \sqrt{|x_{j,k} - x_{j+1,k}|^2 + |x_{j,k} - x_{j,k+1}|^2} \\ &\quad + \sum_{j=1}^{m-1} |x_{j,n} - x_{j+1,n}| + \sum_{k=1}^{n-1} |x_{m,k} - x_{m,k+1}| \\ &= \sum_{j=1}^{m-1}\sum_{k=1}^{n-1} \sqrt{\left(u_{j,k}^{(1)}\right)^2 + \left(v_{j,k}^{(1)}\right)^2 + \left(u_{j,k}^{(2)}\right)^2 + \left(v_{j,k}^{(2)}\right)^2} \\ &\quad + \sum_{j=1}^{m-1} \sqrt{\left(u_{j,n}^{(1)}\right)^2 + \left(v_{j,n}^{(1)}\right)^2} + \sum_{k=1}^{n-1} \sqrt{\left(u_{m,k}^{(2)}\right)^2 + \left(v_{m,k}^{(2)}\right)^2}. \end{aligned} \quad (7)$$

The type-I anisotropic TV is defined as

$$\begin{aligned} \|\mathbf{x}\|_{\text{TV1a}} &= \sum_{j=1}^{m-1}\sum_{k=1}^{n-1} \{|x_{j,k} - x_{j+1,k}| + |x_{j,k} - x_{j,k+1}|\} \\ &\quad + \sum_{j=1}^{m-1} |x_{j,n} - x_{j+1,n}| + \sum_{k=1}^{n-1} |x_{m,k} - x_{m,k+1}| \\ &= \sum_{j=1}^{m-1}\sum_{k=1}^{n-1} \left\{ \sqrt{\left(u_{j,k}^{(1)}\right)^2 + \left(v_{j,k}^{(1)}\right)^2} + \sqrt{\left(u_{j,k}^{(2)}\right)^2 + \left(v_{j,k}^{(2)}\right)^2} \right\} \\ &\quad + \sum_{j=1}^{m-1} \sqrt{\left(u_{j,n}^{(1)}\right)^2 + \left(v_{j,n}^{(1)}\right)^2} + \sum_{k=1}^{n-1} \sqrt{\left(u_{m,k}^{(2)}\right)^2 + \left(v_{m,k}^{(2)}\right)^2}. \end{aligned} \quad (8)$$

Another way to define the complex TV is to separate the real and imaginary parts as two terms, each of which is a real-valued TV. The type-II isotropic TV is defined as

$$\begin{aligned} \|\mathbf{x}\|_{\text{TV2i}} &= \alpha\|\mathcal{R}(\mathbf{x})\|_{\text{TVi}} + (1-\alpha)\|\mathcal{I}(\mathbf{x})\|_{\text{TVi}} \\ &= \alpha\left\{ \sum_{j=1}^{m-1}\sum_{k=1}^{n-1} \sqrt{\left(u_{j,k}^{(1)}\right)^2 + \left(u_{j,k}^{(2)}\right)^2} + \sum_{j=1}^{m-1} |u_{j,n}^{(1)}| + \sum_{k=1}^{n-1} |u_{m,k}^{(2)}| \right\} \\ &\quad + (1-\alpha)\left\{ \sum_{j=1}^{m-1}\sum_{k=1}^{n-1} \sqrt{\left(v_{j,k}^{(1)}\right)^2 + \left(v_{j,k}^{(2)}\right)^2} + \sum_{j=1}^{m-1} |v_{j,n}^{(1)}| + \sum_{k=1}^{n-1} |v_{m,k}^{(2)}| \right\}, \end{aligned} \quad (9)$$

where $\alpha \in [0,1]$ is a parameter. Similarly, the anisotropic TV is given by

$$\|\mathbf{x}\|_{\text{TV2a}} = \alpha \|\mathcal{R}(\mathbf{x})\|_{\text{TVa}} + (1-\alpha)\|\mathcal{I}(\mathbf{x})\|_{\text{TVa}}$$
$$= \alpha \left\{ \sum_{j=1}^{m-1}\sum_{k=1}^{n-1} \left\{ \left|u_{j,k}^{(1)}\right| + \left|u_{j,k}^{(2)}\right| \right\} + \sum_{j=1}^{m-1}\left|u_{j,n}^{(1)}\right| + \sum_{k=1}^{n-1}\left|u_{m,k}^{(2)}\right| \right\}$$
$$+ (1-\alpha) \left\{ \sum_{j=1}^{m-1}\sum_{k=1}^{n-1} \left\{ \left|v_{j,k}^{(1)}\right| + \left|v_{j,k}^{(2)}\right| \right\} + \sum_{j=1}^{m-1}\left|v_{j,n}^{(1)}\right| + \sum_{k=1}^{n-1}\left|v_{m,k}^{(2)}\right| \right\}. \quad (10)$$

In problems (1) and (2), the objective function is a real-valued function over complex-valued variables. This type of problems can be solved either by introducing the CR calculus (i.e., using the Wirtinger derivatives) [27] or by replacing the complex-valued variables by real-valued counterparts. In a recent paper [28], the authors applied the CR calculus and tackled the problem directly in the complex domain, which is beneficial for acoustic wavefield reconstruction. Here we adopt the second approach, since many results from convex analysis can be applied directly in this case. The complex-valued image $\mathbf{x} = \mathbf{u} + i\mathbf{v} \in \mathbb{C}^{m\times n}$ can be equivalently expressed by a pair of real-valued matrices $(\mathbf{u},\mathbf{v}) \in \mathbb{R}^{m\times n} \times \mathbb{R}^{m\times n}$. From now on, $\mathbf{x}$ refers to the matrix pair $(\mathbf{u},\mathbf{v})$ and should be interpreted as a real-valued variable. With this, any function, set, or (semi-)norm of $\mathbf{x}$ defined above should also be interpreted as that of real-valued matrix pairs, but having the same physical meaning and output value as the original one. We further let $\mathcal{T}_{R2C}(\cdot)$ denote the linear operator that transforms a real-valued vector or matrix pair into the original complex-valued vector or matrix. The Frobenius norm of a matrix pair is given by $\|(\mathbf{u},\mathbf{v})\|_F = \sqrt{\|\mathbf{u}\|_F^2 + \|\mathbf{v}\|_F^2}$. To keep notations light, we further let $\mathbf{p}$ denote the group of finite difference matrices $\mathbf{p} = (\mathbf{u}^{(1)},\mathbf{u}^{(2)},\mathbf{v}^{(1)},\mathbf{v}^{(2)}) \in \mathbb{R}^{(m-1)\times n} \times \mathbb{R}^{m\times(n-1)} \times \mathbb{R}^{(m-1)\times n} \times \mathbb{R}^{m\times(n-1)}$.

## III. CONSTRAINED TOTAL VARIATION REGULARIZATION AND THE PROXIMAL GRADIENT METHOD

The optimization problem (2) can be equivalently expressed as

$$\min_{\mathbf{x}} \ F(\mathbf{x}) + \tau \|\mathbf{x}\|_{\text{TV}} + I_C(\mathbf{x}), \quad (11)$$

where $I_C$ is the indicator function of set $C$, whose value is zero if $\mathbf{x} \in C$ and infinity otherwise. In the formulation above, the TV seminorm can be either type-I or type-II, isotropic or anisotropic.

### A. Preliminaries: The Proximal Gradient Method

The inverse problem (11) falls into the category of a more generalized class of optimization problems in the form of

$$\min_{\mathbf{x}} \ F(\mathbf{x}) + R(\mathbf{x}), \quad (12)$$

where $F$ is a smooth function, whereas $R$ is convex but not necessarily differentiable. While the smooth term $F(\mathbf{x})$ can be minimized through a gradient descent scheme, the nonsmooth term $R(\mathbf{x})$ is not readily accessible for gradient calculation,

**Algorithm 1:** Fast Iterative Shrinkage Thresholding Algorithm (FISTA).

**Input:** Initial guess $\mathbf{x}_0$, iteration number $K$, and step size $\gamma$
**Output:** Estimate $\mathbf{x}_K$
1: $\mathbf{z}_1 \leftarrow \mathbf{x}_0, t_1 \leftarrow 1$
2: **for** $k \leftarrow 1$ **to** $K$ **do**
3:     $\mathbf{x}_k \leftarrow \text{prox}_{\gamma R}(\mathbf{z}_k - \gamma \nabla F(\mathbf{y}_k))$
4:     $t_{k+1} \leftarrow \left(1 + \sqrt{1 + 4t_k^2}\right)/2$
5:     $\mathbf{z}_{k+1} \leftarrow \mathbf{x}_k + \left(\frac{t_k - 1}{t_{k+1}}\right)(\mathbf{x}_k - \mathbf{x}_{k-1})$
6: **end**

and is thus accessed only via its proximity operator, which is defined as

$$\text{prox}_{\gamma R}(\mathbf{x}) = \arg\min_{\mathbf{z}} \left\{ R(\mathbf{z}) + \frac{1}{2\gamma}\|\mathbf{z} - \mathbf{x}\|_F^2 \right\}, \quad (13)$$

where $\gamma > 0$ is a parameter. The proximity operator can be interpreted as a generalization of the gradient descent method to nonsmooth functions with respect to the Moreau envelope. The proximal gradient method is a classical approach to the nonsmooth minimization problem (12), whose simplest form can be expressed as [22]

$$\mathbf{x}_{k+1} = \text{prox}_{\gamma_k R}\left(\mathbf{x}_k - \gamma_k \nabla F(\mathbf{x}_k)\right), \quad (14)$$

where the subscript denotes the iteration number. Within each iteration, the algorithm updates $F$ and $R$ in an alternative manner, which requires calculating of the gradient and the proximity operator, respectively. As a result, the proximal gradient method is particularly favorable when the gradient of $F$ with respect to $\mathbf{x}$ can be expressed explicitly and the subproblem (13) can be solved efficiently.

### B. Accelerated Version: FISTA

There are many accelerated variants of the basic proximal gradient algorithm [29]–[31]. In this work, we adopt the fast iterative shrinkage / thresholding algorithm (FISTA) [31], which has been successfully applied to many inverse problems. Although it was initially proposed to solve convex optimization problems, our numerical studies have shown that the nonconvex version of FISTA also converges much faster than the basic proximal gradient algorithm known as the iterative shrinkage / thresholding algorithm (ISTA) [32]–[34]. The pseudocode for FISTA is presented in Algorithm 1. In our numerical studies, the step size $\gamma$ is selected via a line search strategy.

## IV. CONSTRAINED TOTAL VARIATION DENOISING

In this Section, we derive an efficient algorithm for solving the subproblem (13) with $R(\mathbf{x}) = \tau \|\mathbf{x}\|_{\text{TV}} + I_C(\mathbf{x})$, which can be plugged into the FISTA framework for solving the general problem (11). The subproblem (13) can be reformulated as a constrained total variation denoising problem:

$$\min_{\mathbf{x}} \ \frac{1}{2}\|\mathbf{x} - \mathbf{b}\|_F^2 + \underbrace{\lambda \|\mathbf{x}\|_{\text{TV}} + I_C(\mathbf{x})}_{\gamma R(\mathbf{x})}, \quad (15)$$

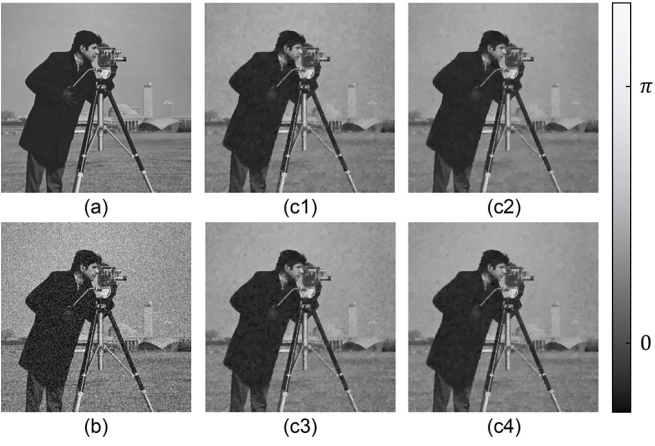

Fig. 1. Denoising results with respect to the argument of a complex-valued image. Since the modulus of the original image is assumed to be uniform, only the complex arguments are presented. (a) Original image. (b) Noisy image. (c1-c4) Denoised image with the (c1) type-I isotropic, (c2) type-I anisotropic, (c3) type-II isotropic, and (c4) type-II anisotropic TV.

where $\mathbf{b} \in \mathbb{R}^{m \times n} \times \mathbb{R}^{m \times n}$, and $\lambda = \tau \gamma$ is a parameter. It is self-evident that (15) is equivalent to (1). Therefore, the constrained denoising problem can be indeed regarded as a proximal update in the proximal gradient method.

### A. Dual Formulation

The main difficulty of solving problem (15) comes from the nonsmoothness of the TV seminorm and the indicator function. In [23] and [24], the authors considered tackling the primal problem via its dual, which turns out to be much easier to solve. Here, we follow their approach and derive the dual problem. For the sake of simplicity, we take the type-I isotropic TV as an example in the following derivation.

*Proposition 1:* The dual problem of (14) is given by

$$\min_{\mathbf{q} \in S} \left\{ h(\mathbf{q}) \equiv -\left\| \mathcal{H}_C \left( \mathbf{b} - \lambda \mathcal{L}^T (\mathbf{q}) \right) \right\|_F^2 + \left\| \mathbf{b} - \lambda \mathcal{L}^T (\mathbf{q}) \right\|_F^2 \right\}, \quad (16)$$

where $\mathcal{H}_C(\cdot)$ is a linear operator such that $\mathcal{H}_C(\mathbf{x}) = \mathbf{x} - \mathcal{P}_C(\mathbf{x})$ and $\mathcal{P}_C(\cdot)$ is the projection operator of set $C$. $\mathcal{L}(\cdot)$ is the linear map from the image $\mathbf{x}$ into its finite differences $\mathbf{p}$, and $\mathcal{L}^T(\cdot)$ denotes the adjoint operator of $\mathcal{L}(\cdot)$. $\mathbf{q} = \left(\mathbf{r}^{(1)}, \mathbf{r}^{(2)}, \mathbf{s}^{(1)}, \mathbf{s}^{(2)}\right) \in \mathbb{R}^{(m-1) \times n} \times \mathbb{R}^{m \times (n-1)} \times \mathbb{R}^{(m-1) \times n} \times \mathbb{R}^{m \times (n-1)}$ is the dual variable. $S$ is the set of all matrix groups $\mathbf{q} = \left(\mathbf{r}^{(1)}, \mathbf{r}^{(2)}, \mathbf{s}^{(1)}, \mathbf{s}^{(2)}\right)$ that satisfy

$$\begin{aligned}
&\left(r_{j,k}^{(1)}\right)^2 + \left(r_{j,k}^{(2)}\right)^2 + \left(s_{j,k}^{(1)}\right)^2 + \left(s_{j,k}^{(2)}\right)^2 \leq 1, \\
&\qquad\qquad j = 1,\ldots,m-1, k = 1,\ldots,n-1 \\
&\left(r_{j,n}^{(1)}\right)^2 + \left(s_{j,n}^{(1)}\right)^2 \leq 1, \qquad j = 1,\ldots,m-1 \\
&\left(r_{m,k}^{(2)}\right)^2 + \left(s_{m,k}^{(2)}\right)^2 \leq 1, \qquad k = 1,\ldots,n-1.
\end{aligned} \quad (17)$$

In addition, the primal optimal solution $\mathbf{x}^\star$ is related to the dual optimal solution $\mathbf{q}^\star$ via

$$\mathbf{x}^\star = \mathcal{P}_C \left( \mathbf{b} - \lambda \mathcal{L}^T (\mathbf{q}^\star) \right). \quad (18)$$

Therefore, to solve the primal problem, we only need to solve the dual problem, which is made much easier with a gradient projection algorithm described in the next Section. Then the primal solution is directly obtained using Eq. (18). For the other three types of TV, the above proposition holds, except that the set $S$ is defined differently. See Appendix B for details.

### B. The Gradient Projection Algorithm

In this Section, we present a gradient projection algorithm that solves the dual problem efficiently. The algorithm proceeds by iteratively update the variable $\mathbf{q}$ with a gradient descent scheme combined with a projection operation. Although the following results are mainly obtained from [24], we still present them here for completeness.

The objective function $h(\mathbf{q})$ of (16) has been proved to be continuously differentiable, and the gradient is given by

$$\nabla h(\mathbf{q}) = -2\lambda \mathcal{L} \left( \mathcal{P}_C \left( \mathbf{b} - \lambda \mathcal{L}^T (\mathbf{q}) \right) \right). \quad (19)$$

Furthermore, following the proof in [24], one can easily verify that $\nabla h(\mathbf{q})$ has a Lipschitz upper bound $16\lambda^2$. The constraint $\mathbf{q} \in S$ is enforced via projections onto the set $S$. For the type-I isotropic TV, the projection $\tilde{\mathbf{q}} = \mathcal{P}_S(\mathbf{q}) = \left(\tilde{\mathbf{r}}^{(1)}, \tilde{\mathbf{r}}^{(2)}, \tilde{\mathbf{s}}^{(1)}, \tilde{\mathbf{s}}^{(2)}\right)$ is given by

$$\tilde{\eta}_{j,k}^{(1)} = \begin{cases} \dfrac{\eta_{j,k}^{(1)}}{\max\left\{1, \sqrt{\left(r_{j,k}^{(1)}\right)^2 + \left(r_{j,k}^{(2)}\right)^2 + \left(s_{j,k}^{(1)}\right)^2 + \left(s_{j,k}^{(2)}\right)^2}\right\}}, \\ \qquad j = 1,\ldots,m-1, k = 1,\ldots,n-1 \\ \dfrac{\eta_{j,k}^{(1)}}{\max\left\{1, \sqrt{\left(r_{j,k}^{(1)}\right)^2 + \left(s_{j,k}^{(1)}\right)^2}\right\}}, \\ \qquad j = 1,\ldots,m-1, k = n \end{cases}$$

$$\tilde{\eta}_{j,k}^{(2)} = \begin{cases} \dfrac{\eta_{j,k}^{(2)}}{\max\left\{1, \sqrt{\left(r_{j,k}^{(1)}\right)^2 + \left(r_{j,k}^{(2)}\right)^2 + \left(s_{j,k}^{(1)}\right)^2 + \left(s_{j,k}^{(2)}\right)^2}\right\}}, \\ \qquad j = 1,\ldots,m-1, k = 1,\ldots,n-1 \\ \dfrac{\eta_{j,k}^{(2)}}{\max\left\{1, \sqrt{\left(r_{j,k}^{(2)}\right)^2 + \left(s_{j,k}^{(2)}\right)^2}\right\}}, \\ \qquad j = m, k = 1,\ldots,n-1 \end{cases} \quad (20)$$

---

**Algorithm 2:** Fast Gradient Projection (FGP) Algorithm.

**Input:** Observation $\mathbf{b}$, iteration number $K$, and parameter $\lambda$
**Output:** Estimate $\mathbf{x}$

1: $\mathbf{q}_0 \leftarrow \left(\mathbf{0}_{(m-1) \times n}, \mathbf{0}_{m \times (n-1)}, \mathbf{0}_{(m-1) \times n}, \mathbf{0}_{m \times (n-1)}\right)$,
   $\mathbf{r}_1 \leftarrow \mathbf{q}_0, t_1 \leftarrow 1$
2: **for** $k \leftarrow 1$ **to** $K$ **do**
3: $\quad \mathbf{q}_k \leftarrow \mathcal{P}_S\left(\mathbf{r}_k - \frac{1}{16\lambda^2} \nabla h(\mathbf{r}_k)\right)$
4: $\quad t_{k+1} \leftarrow \left(1 + \sqrt{1 + 4t_k^2}\right)/2$
5: $\quad \mathbf{r}_{k+1} \leftarrow \mathbf{q}_k + \left(\frac{t_k - 1}{t_{k+1}}\right)(\mathbf{q}_k - \mathbf{q}_{k-1})$
6: **end**

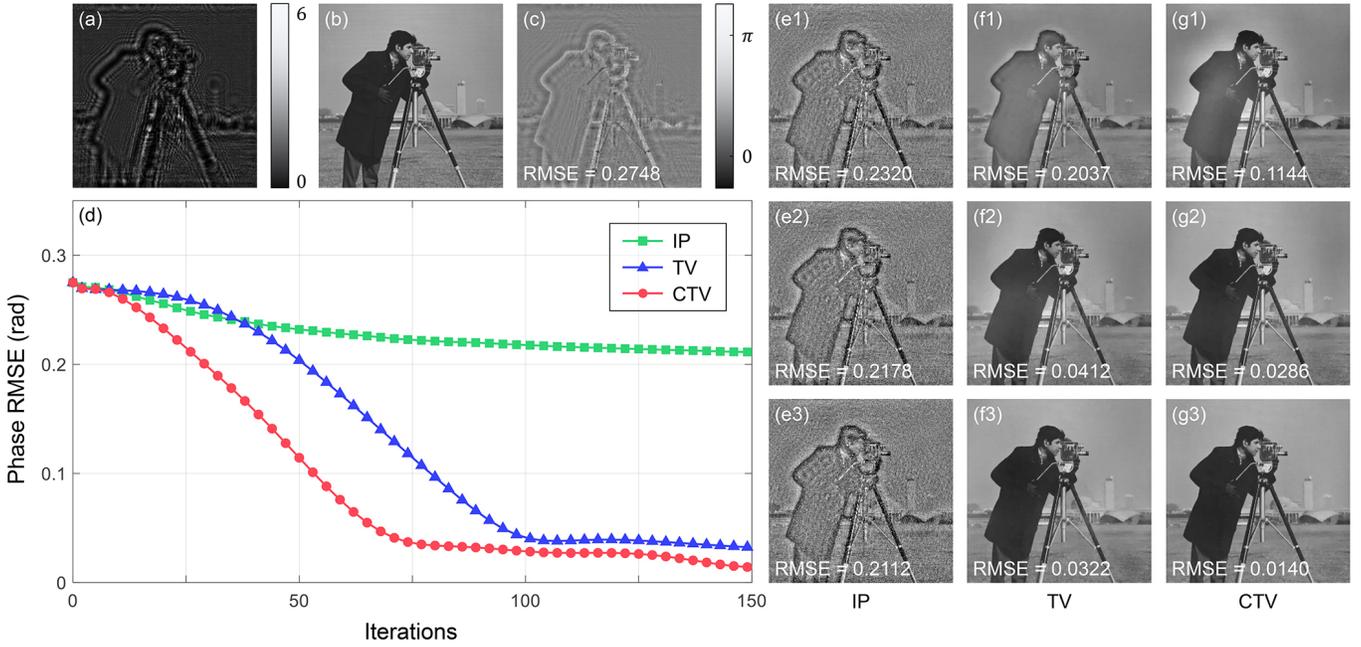

Fig. 2. Comparison of the phase retrieval results using different methods. (a) The real-valued intensity measurement **y**. (b) Phase of the virtual object. (c) Phase of the directly back-propagated complex field. (d) RMSEs with respect to the phase reconstruction. (e1-e3) Recovered phase using the IP method at the 50th, 100th, and 150th iteration, respectively. (f1-f3) Recovered phase using the TV method at the 50th, 100th, and 150th iteration, respectively. (g1-g3) Recovered phase using the CTV method at the 50th, 100th, and 150th iteration, respectively.

where $\eta \in \{r,s\}$. The projection operators for the other three cases are given in Appendix B. The basic gradient projection algorithm can be expressed as

$$\mathbf{q}_{k+1} = \mathcal{P}_S\left(\mathbf{q}_k - \frac{1}{16\lambda^2}\nabla h(\mathbf{q}_k)\right). \quad (21)$$

Similar to the idea of FISTA, an accelerated version, termed the fast gradient projection (FGP) algorithm, has been proposed in [24] for faster convergence, and is then generalized by us to the complex-valued case, which is described in Algorithm 2.

### C. Numerical Examples on Denoising

We tested the denoising algorithm with a 256×256 complex-valued image. The image has an all-ones uniform modulus and an argument of the cameraman image, which is shown in Fig. 1. A normally distributed noise with a standard deviation of $\pi/10$ was added to the argument of each pixel.

As a visualized demonstration, the denoised arguments are shown in Fig. 1 (c1-c4) using the four types of complex TVs. For the type-I TVs, we set the regularization parameter $\lambda = 0.2$. For the type-II TVs, we set $\lambda = 0.3$, and the regularization weight $\alpha = 1/2$. In all four cases, we ran the FGP algorithm for 50 iterations. We observed that with a fixed $\lambda$, the type-I TVs usually impose a stronger regularization effect than the type-II TVs do, and the smoothing effect of the anisotropic TVs is stronger than that of the isotropic ones. But all four TVs can achieve similar denoising effects by tuning $\lambda$.

## V. APPLICATION TO PHASE RETRIEVAL

### A. Forward Model

Phase retrieval is a classical problem well-known to the optics and signal processing community. It forms the basis of many imaging applications ranging from digital holographic imaging to X-ray crystallography. For high-frequency electromagnetic waves, sensors can only respond to the average power of the incident wavefield, thus the phase information, i.e., the complex argument of the signal, is lost during the measurement process. The general forward model can be expressed as

$$|\mathcal{A}(\mathbf{x})|^2 = \mathbf{y}, \quad (22)$$

where $\mathcal{A}(\cdot)$ denotes the linear forward transmission operator which is determined by the imaging system, **x** is the real-valued matrix pairs that represent the complex transmittance of the object, and **y** is the corresponding intensity measurement. The forward model (25) holds so long as the illumination source is coherent, the propagation of the wavefield satisfies the Scalar Diffraction Theory [17], and the wavefield is adequately sampled by the pixelated sensor [35]. In this paper, we consider the simplest Gabor holography [36] as an example, where the linear operator represents a free-space propagation:

$$\mathcal{A}(\cdot) = \mathcal{F}^{-1}\left\{\mathcal{F}[\mathcal{T}_{R2C}(\cdot)]\exp\left(i\frac{2\pi}{\lambda_s}d\sqrt{1-(\lambda_s f_x)^2-(\lambda_s f_y)^2}\right)\right\}, \quad (23)$$

where $\lambda_s$ is the wavelength of the illumination source, $d$ is the propagation distance, $f_x$ and $f_y$ are the spatial frequency coordinates, $\mathcal{F}(\cdot)$ and $\mathcal{F}^{-1}(\cdot)$ denote the Fourier transform and inverse Fourier transform, respectively.

### B. Problem Formulation

Recovering a complex signal from real-valued measurements is generally difficult. In particular, when there is only a single intensity image, the problem is well-known to be ill-posed [26]. For reconstruction purposes, the model above is reformulated as a regularized optimization problem:

$$\min_{\mathbf{x}} \underbrace{1/2\|\|\mathcal{A}(\mathbf{x})\|-\sqrt{\mathbf{y}}\|_F^2}_{F(\mathbf{x})} + \underbrace{\tau\|\mathbf{x}\|_{\mathrm{TV}} + I_C(\mathbf{x})}_{R(\mathbf{x})}. \quad (24)$$

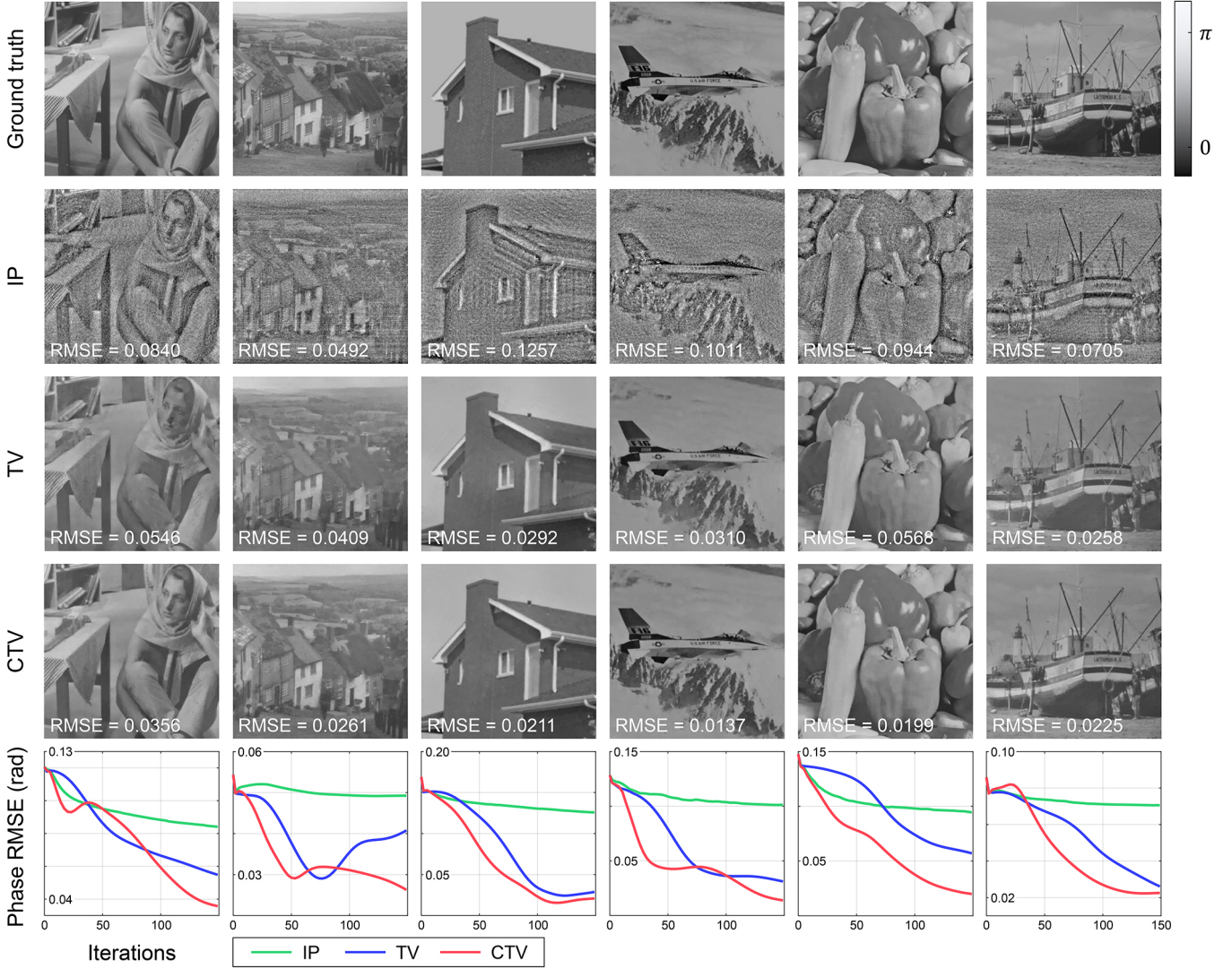

Fig. 3. Comparison of the phase retrieval results with different simulated objects using the IP, TV, and CTV methods.

Therefore, the phase retrieval problem (24) can be viewed as a special case of (11). $F(\mathbf{x})$ is the data-fidelity term that is minimized when the current estimate is consistent with the measurement. It should be noted that, although strictly speaking, the fidelity term $F(\mathbf{x})$ is not everywhere differentiable, its non-smoothness can be easily addressed. See the next Section for details. $R(\mathbf{x})$ serves as the regularization function, which has two separate terms with different physical meanings.

The introduction of the indicator function $I_C$ is inspired by the conventional phase retrieval methods. They assume that the complex object satisfies certain prior constraints, which can be expressed in the form of a set $C$ [37]–[39]. For example, based on the notion of energy conservation, the unknown object is often assumed to have nonnegative absorption values [40]. The corresponding constraint set is given by

$$C = \{\mathbf{x} = (\mathbf{u}, \mathbf{v}) \in \mathbb{R}^{m \times n} \times \mathbb{R}^{m \times n} : \mathbf{u}^2 + \mathbf{v}^2 \leq \mathbf{1}\}, \quad (25)$$

which is a closed convex set for $\mathbf{x} = (\mathbf{u}, \mathbf{v})$.

The TV seminorm serves as a sparsity-promoting regularizer in (24). It should be noted that, we are not the first to exploit sparsity priors in phase retrieval. Early works proposed using $l_1$ regularization to enforce sparsity [41], [42]. Later in [43], the authors introduced the TV seminorm as a more powerful tool for holographic reconstruction. Nevertheless, the method of [43] is limited in the following aspects. First, the derivation was based on the conventional real-valued TV seminorm, and thus cannot be extended to arbitrary complex objects. Second, the forward model was linearized by dropping the higher-order terms. This makes solving the problem much easier using the state-of-the-art algorithms for linear inverse problems, but it sacrifices the reconstruction fidelity and may result in a large error when the object violates the weak object approximation. Third, the previous work did not consider any constraint sets that had been well studied before for phase retrieval, which could potentially help improve the reconstruction quality or the rate of convergence.

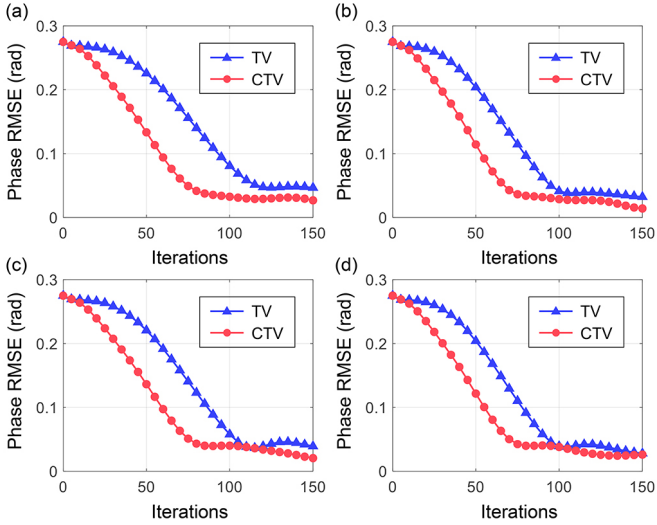

Fig. 4. Comparison of the TV and CTV models with respect to the phase RMSEs using (a) type-I isotropic TV, (b) type-I anisotropic TV, (c) type-II isotropic TV, and (d) type-II anisotropic TV seminorms.

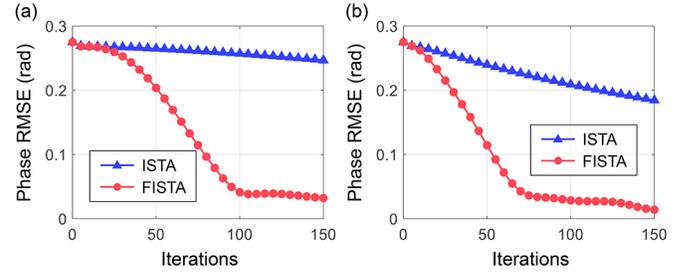

Fig. 5. Comparison of the convergence behavior of ISTA and FISTA using the (a) TV and (b) CTV models.

### C. Derivation of the Algorithm

To apply FISTA to phase retrieval, we need to provide the analytical expressions necessary for numerical implementation. As mentioned above, the non-smooth term $R(\mathbf{x})$ is updated via the proximity operator, which is solved by the FGP algorithm. The corresponding projection $(\tilde{\mathbf{u}}, \tilde{\mathbf{v}}) = \mathcal{P}_C(\mathbf{u}, \mathbf{v})$ is given by

$$\tilde{u}_{j,k} = \frac{u_{j,k}}{\max\left\{1, \sqrt{u_{j,k}^2 + v_{j,k}^2}\right\}} \quad j=1,\ldots,m, k=1,\ldots,n$$
$$\tilde{v}_{j,k} = \frac{v_{j,k}}{\max\left\{1, \sqrt{u_{j,k}^2 + v_{j,k}^2}\right\}} \quad j=1,\ldots,m, k=1,\ldots,n. \tag{26}$$

The smooth term $F(\mathbf{x})$ is updated via its gradient, which can be expressed as

$$\nabla F(\mathbf{x}) = \mathcal{R}\left\{\mathcal{A}^*\left[\left(\frac{1}{|\mathcal{A}(\mathbf{x})|} \odot \mathcal{A}(\mathbf{x})\right) \odot \left(|\mathcal{A}(\mathbf{x})| - \sqrt{\mathbf{y}}\right)\right]\right\}. \tag{27}$$

See Appendix C for derivation. With this, we now obtain the FISTA for phase retrieval. The MATLAB code can be found in https://github.com/THUHoloLab/complex-constrained-total-variation-denoising.

## VI. NUMERICAL EXAMPLES ON PHASE RETRIEVAL

Our proposed TV model and the algorithmic framework have been corroborated with numerical tests based on both simulated and experimental data. We mainly consider pure phase objects because they contain nonzero real and imaginary parts, and are frequently encountered in imaging problems. All the numerical experiments presented in this paper were implemented with MATLAB R2019a on a laptop computer with Core i5-10210U CPU @ 1.6GHz (Intel) and 16 GB of RAM.

### A. Simulated Experiments

The simulated pure phase objects were generated using the standard test images, with a phase distribution ranging from 0 to $\pi$ and a size of 256×256 square pixels. In all experiments, the pixel size was 5μm, the illumination wavelength was 500nm, and the imaging distance was 5mm. Numerical propagation of the wavefield was implemented using the angular spectrum method [44]. 10% Gaussian noise was added to the calculated intensity, formulating the measurement from the sensor, which is shown in Fig. 2(a). All the tested algorithms were initialized with the directly back-propagated complex wavefield. The resulting initial guess suffers from a severe twin-image artifact, because of the missing phase at the sensor plane. Root-mean-square errors (RMSEs) with respect to the recovered phase were calculated to quantify the reconstruction errors. It should be noted that, since we are interested in the relative phase of the object, a global phase shift was not considered as an error in our experiments.

The proposed model is evaluated relative to the conventional iterative projection (IP) method [40]. The IP algorithm proceeds by alternatively projecting the current estimate of the complex object onto the modulus constraint set and the nonnegative absorption constraint set. For further illustration, we tested the TV model for both constrained and unconstrained cases. In the constrained case, the set $C$ is defined by (25), whereas in the unconstrained case, we simply set $C = \mathbb{C}^{m \times n}$. The former is referred to as the constrained TV (CTV) method, whereas the latter is simply termed the TV method. For both TV and CTV algorithms, we ran the FGP algorithm for 10 iterations to solve the denoising subproblem within each main loop.

Figure 2 visualizes the convergence behavior of the above three methods. The results were obtained using the type-I anisotropic TV regularizer unless otherwise stated. The TV-based methods outperform the conventional IP method in terms of reconstruction quality and convergence rate. The results of the TV and CTV methods are similar, but with the additional physical constraint, the CTV method converges faster at the beginning stage. One can also observe that the low-frequency part of the image is more difficult to recover than its sharp edges. This agrees with the general notion in Fourier optics that the low-frequency phase can hardly contribute to the diffraction intensity [45].

To study the scalability and efficacy of the proposed complex TV model on natural images, we ran the phase retrieval algorithms for reconstruction of another six simulated pure phase objects, as is shown in Fig. 3. We discovered that the influence of the constraint set is twofold. It helps improve the robustness of the algorithm against various settings, yielding much more robust convergence behavior when compared with the unconstrained TV method. On the other hand, the additional physical boundary introduced by the constraint may limit the search space for the estimate, which could possibly slow down the convergence. Nevertheless, the latter seems less of a problem from the numerical results presented. We also tested the TV and CTV methods using different TV seminorms. The results, which are presented in Fig. 4, mostly agree with the above findings.

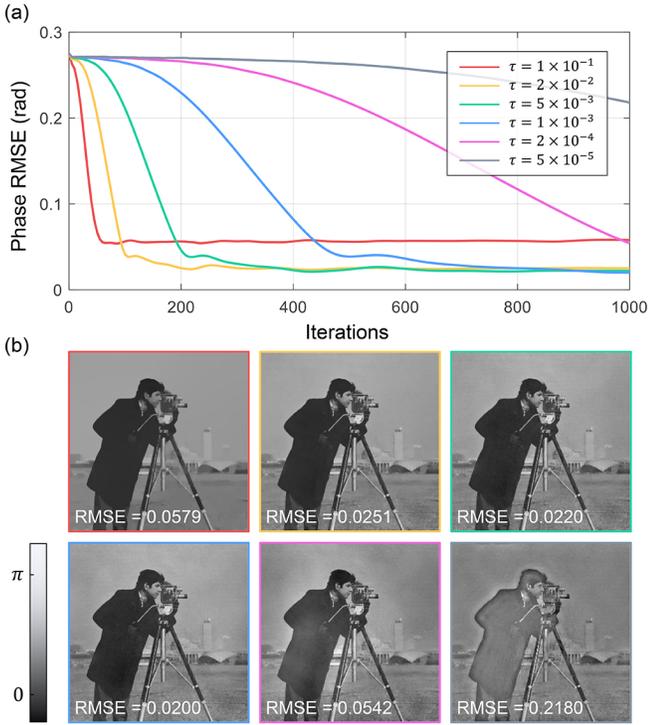

Fig. 6. Comparison of different choices of the regularization parameter $\tau$. (a) The phase RMSE curves. (b) Reconstructed phase with different choices of $\tau$ after 1000 iterations, with the phase RMSE given below.

We compared FISTA with the basic proximal gradient algorithm, which is known as ISTA. The results are shown in Fig. 5. Evidently, the accelerated FISTA converges much faster than ISTA, although the former was not initially proposed for solving nonconvex optimization problems.

To examine how the regularization parameter $\tau$ influences the reconstruction results, we ran the CTV algorithm with different values of $\tau$. The loss curves and visualized reconstruction are given in Fig. 6. Imposing a stronger regularization can speed up the algorithm at the beginning stage. But there exists an optimal value of $\tau$ in terms of the reconstruction fidelity or algorithm efficiency. From the visualized reconstruction, we can see that the selection of the regularization parameter faces a balance between the loss of details and the background noise.

### B. Optical Experiments

We have conducted optical experiments to verify our model in real-world settings. A Quantitative Phase Microscopy Target (Benchmark Technologies) was illuminated by a plane wave at the wavelength of 660nm. A CCD imaging sensor with a pixel size of 5.74μm was placed 6.7mm behind the sample.

For comparison, we also implemented a conventional multi-image phase retrieval method. The observation diversity was achieved by using a programmable spatial light modulator. Readers may refer to [46] for a more detailed description of the system configuration. In total eight images were captured with different phase modulation patterns. We then ran the Wirtinger gradient descent algorithm for reconstruction [35].

The experiment results are presented in Fig. 7, where the conventional method is compared to the proposed CTV method. For high-frequency parts, the phase can be accurately recovered using both methods, with the CTV method sacrificing a little bit of resolution. For the low-frequency parts, the CTV method outperforms the conventional method in recovering the phase. As mentioned before, this is because of the inherent limitation of the inline configuration in recording low-frequency phase information. Furthermore, recovering the low-frequency phase from intensity measurements can be particularly difficult in the existence of noise and system miscalibrations. Nevertheless, this problem can be in part addressed using the sparsity priors. Most importantly, the CTV method only requires one intensity image for reconstruction, in sharp contrast to the conventional multi-image approach.

### VII. CONCLUSION

We introduced two types of TV seminorms for complex-valued images in both isotropic and anisotropic forms. We proposed a fast gradient projection algorithm for solving the complex constrained TV image denoising problem. The proposed algorithm can serve as a key building block for a vast range of complex constrained minimization problems with TV-based regularizers. We further demonstrated the effectiveness of the proposed algorithmic framework in the context of phase retrieval. Based on numerical studies on both simulated and experimental data, we observed that the proposed constrained TV model shares a fast rate of convergence together with high reconstruction fidelity. Further issues regarding the model, such as the introduction of other sparsity priors or application to other imaging problems, remain to be investigated in the future.

### APPENDIX A
### PROOF OF PROPOSITION 1

We first reformulate the unconstrained problem (15) as an equivalent constrained one:

$$\min_{\mathbf{x},\mathbf{p}} \quad \frac{1}{2\lambda}\|\mathbf{x}-\mathbf{b}\|_F^2 + I_C(\mathbf{x}) + f(\mathbf{p}) \quad (28)$$
$$\text{s.t.} \quad \mathbf{p} = \mathcal{L}(\mathbf{x})$$

where $f(\mathbf{p}) = \|\mathbf{x}\|_{\text{TV1i}}$. The Lagrangian is given by

$$L(\mathbf{x},\mathbf{p},\mathbf{q}) = \frac{1}{2\lambda}\|\mathbf{x}-\mathbf{b}\|_F^2 + I_C(\mathbf{x}) + f(\mathbf{p}) + \langle \mathbf{q}, \mathcal{L}(\mathbf{x}) - \mathbf{p} \rangle \quad (29)$$

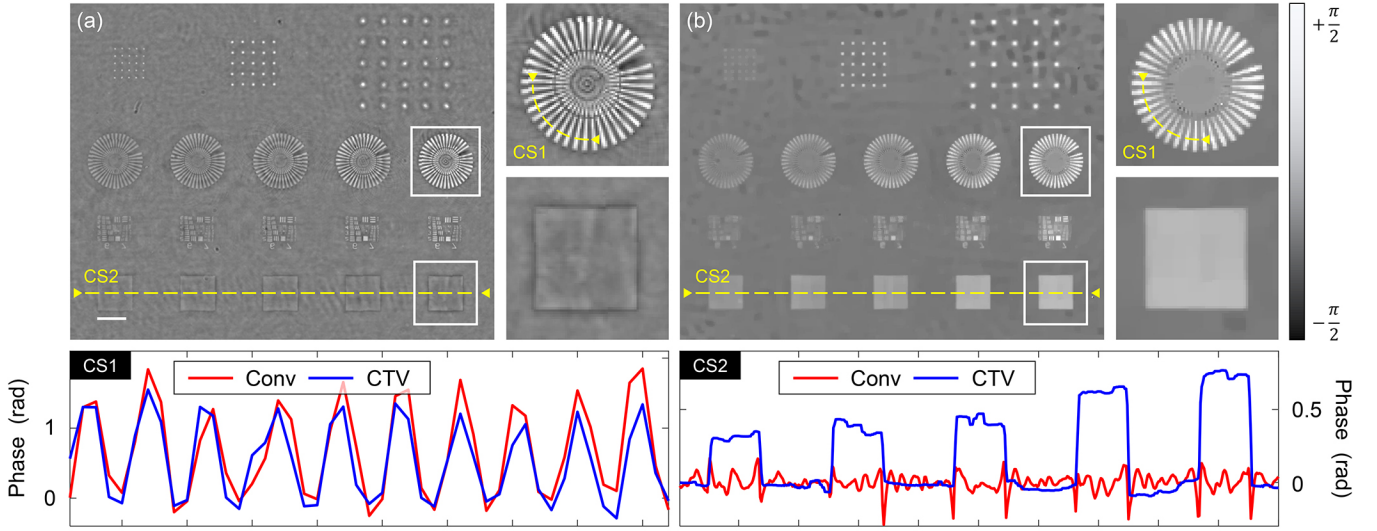

Fig. 7. Comparison of the CTV method against conventional multi-image phase retrieval method (Conv) based on experimental data. (a) Recovered phase using the conventional method. (b) Recovered phase using the CTV method. CS1 and CS2 are the cross-sectional profiles extracted from the original phase images. The white scale bar in (a) is 200μm.

where $\mathbf{q} \in \mathbb{R}^{(m-1)\times n} \times \mathbb{R}^{m\times(n-1)} \times \mathbb{R}^{(m-1)\times n} \times \mathbb{R}^{m\times(n-1)}$ is the dual variable. The Lagrange dual function, by definition, is

$$\inf_{\mathbf{x},\mathbf{p}} L(\mathbf{x},\mathbf{p},\mathbf{q})$$

$$= \inf_{\mathbf{x},\mathbf{p}} \left\{ \frac{1}{2\lambda}\|\mathbf{x}-\mathbf{b}\|_F^2 + I_C(\mathbf{x}) + f(\mathbf{p}) + \langle \mathbf{q}, \mathcal{L}(\mathbf{x}) - \mathbf{p}\rangle \right\}$$

$$= \inf_{\mathbf{x}} \left\{ \frac{1}{2\lambda}\|\mathbf{x}-\mathbf{b}\|_F^2 + I_C(\mathbf{x}) + \langle \mathbf{q}, \mathcal{L}(\mathbf{x})\rangle \right\}$$
$$+ \inf_{\mathbf{p}} \left\{ f(\mathbf{p}) + \langle \mathbf{q}, -\mathbf{p}\rangle \right\}$$

$$= \inf_{\mathbf{x}\in C}\left\{\frac{1}{2\lambda}\|\mathbf{x}-\mathbf{b}\|_F^2 + \langle \mathbf{q},\mathcal{L}(\mathbf{x})\rangle\right\} - \sup_{\mathbf{p}}\left\{\langle \mathbf{q},\mathbf{p}\rangle - f(\mathbf{p})\right\} \quad (30)$$

$$= \inf_{\mathbf{x}\in C}\left\{\frac{1}{2\lambda}\|\mathbf{x}-\mathbf{b}\|_F^2 + \langle \mathcal{L}^T(\mathbf{q}),\mathbf{x}\rangle\right\} - f^*(\mathbf{q})$$

$$= \inf_{\mathbf{x}\in C}\left\{\frac{1}{2\lambda}\|\mathbf{x}-(\mathbf{b}-\lambda\mathcal{L}^T(\mathbf{q}))\|_F^2\right\}$$
$$+ \frac{1}{2\lambda}\|\mathbf{b}\|_F^2 - \frac{1}{2\lambda}\|\mathbf{b}-\lambda\mathcal{L}^T(\mathbf{q})\|_F^2 - I_S(\mathbf{q}),$$

where $f^*$ is the convex conjugate of $f$, which is in fact the indicator function of the set $S$ [47]. The last equality holds because the infimum is obtained when $\mathbf{x} = \mathcal{P}_C(\mathbf{b}-\lambda\mathcal{L}^T(\mathbf{q}))$. By dropping out the constant term, the dual problem can be equivalently expressed as

$$\max_{\mathbf{q}\in S}\left\{\|\mathcal{P}_C(\mathbf{b}-\lambda\mathcal{L}^T(\mathbf{q}))-(\mathbf{b}-\lambda\mathcal{L}^T(\mathbf{q}))\|_F^2 - \|\mathbf{b}-\lambda\mathcal{L}^T(\mathbf{q})\|_F^2\right\}, \quad (31)$$

which is equivalent to (16).

## APPENDIX B
### DEFINITION OF $S$ AND THE PROJECTION OPERATORS

For the type-I anisotropic TV, the set $S$ contains all matrix groups $\mathbf{q}$ that satisfy

$$\begin{aligned}\left(r_{j,k}^{(1)}\right)^2 + \left(s_{j,k}^{(1)}\right)^2 &\leq 1, \quad j=1,\ldots,m-1, k=1,\ldots,n \\ \left(r_{j,k}^{(2)}\right)^2 + \left(s_{j,k}^{(2)}\right)^2 &\leq 1, \quad j=1,\ldots,m, k=1,\ldots,n-1.\end{aligned} \quad (32)$$

The projection operator with respect to $S$, which is denoted as $\tilde{\mathbf{q}} = \mathcal{P}_S(\mathbf{q}) = (\tilde{\mathbf{r}}^{(1)},\tilde{\mathbf{r}}^{(2)},\tilde{\mathbf{s}}^{(1)},\tilde{\mathbf{s}}^{(2)})$, is given by

$$\begin{aligned}\tilde{\eta}_{j,k}^{(1)} &= \frac{\eta_{j,k}^{(1)}}{\max\left\{1,\sqrt{\left(r_{j,k}^{(1)}\right)^2+\left(s_{j,k}^{(1)}\right)^2}\right\}}, \quad j=1,\ldots,m-1, k=1,\ldots,n \\ \tilde{\eta}_{j,k}^{(2)} &= \frac{\eta_{j,k}^{(2)}}{\max\left\{1,\sqrt{\left(r_{j,k}^{(2)}\right)^2+\left(s_{j,k}^{(2)}\right)^2}\right\}}, \quad j=1,\ldots,m, k=1,\ldots,n-1\end{aligned} \quad (33)$$

where $\eta \in \{r,s\}$.

For the type-II isotropic TV, $S$ consists of all $\mathbf{q}$ satisfying

$$\begin{aligned}\left(r_{j,k}^{(1)}\right)^2 + \left(r_{j,k}^{(2)}\right)^2 &\leq \alpha^2, \quad j=1,\ldots,m-1, k=1,\ldots,n-1 \\ \left|r_{j,n}^{(1)}\right| &\leq \alpha, \quad j=1,\ldots,m \\ \left|r_{m,k}^{(2)}\right| &\leq \alpha, \quad k=1,\ldots,n \\ \left(s_{j,k}^{(1)}\right)^2 + \left(s_{j,k}^{(2)}\right)^2 &\leq (1-\alpha)^2, \quad j=1,\ldots,m-1, k=1,\ldots,n-1 \\ \left|s_{j,n}^{(1)}\right| &\leq 1-\alpha, \quad j=1,\ldots,m \\ \left|s_{m,k}^{(2)}\right| &\leq 1-\alpha, \quad k=1,\ldots,n.\end{aligned} \quad (34)$$

The corresponding projection is given by

$$\tilde{r}_{j,k}^{(1)} = \begin{cases} \dfrac{\alpha r_{j,k}^{(1)}}{\max\left\{\alpha, \sqrt{\left(r_{j,k}^{(1)}\right)^2 + \left(r_{j,k}^{(2)}\right)^2}\right\}}, & \begin{array}{l} j=1,\ldots,m-1, \\ k=1,\ldots,n-1 \end{array} \\[2ex] \dfrac{\alpha r_{j,k}^{(1)}}{\max\left\{\alpha, \left|r_{j,k}^{(1)}\right|\right\}}, & \begin{array}{l} j=1,\ldots,m-1, \\ k=n \end{array} \end{cases}$$

$$\tilde{r}_{j,k}^{(2)} = \begin{cases} \dfrac{\alpha r_{j,k}^{(2)}}{\max\left\{\alpha, \sqrt{\left(r_{j,k}^{(1)}\right)^2 + \left(r_{j,k}^{(2)}\right)^2}\right\}}, & \begin{array}{l} j=1,\ldots,m-1, \\ k=1,\ldots,n-1 \end{array} \\[2ex] \dfrac{\alpha r_{j,k}^{(2)}}{\max\left\{\alpha, \left|r_{j,k}^{(2)}\right|\right\}}, & \begin{array}{l} j=m, \\ k=1,\ldots,n-1 \end{array} \end{cases}$$

$$\tilde{s}_{j,k}^{(1)} = \begin{cases} \dfrac{(1-\alpha) s_{j,k}^{(1)}}{\max\left\{1-\alpha, \sqrt{\left(s_{j,k}^{(1)}\right)^2 + \left(s_{j,k}^{(2)}\right)^2}\right\}}, & \begin{array}{l} j=1,\ldots,m-1, \\ k=1,\ldots,n-1 \end{array} \\[2ex] \dfrac{(1-\alpha) s_{j,k}^{(1)}}{\max\left\{1-\alpha, \left|s_{j,k}^{(1)}\right|\right\}}, & \begin{array}{l} j=1,\ldots,m-1, \\ k=n \end{array} \end{cases}$$

$$\tilde{s}_{j,k}^{(2)} = \begin{cases} \dfrac{(1-\alpha) s_{j,k}^{(2)}}{\max\left\{1-\alpha, \sqrt{\left(s_{j,k}^{(1)}\right)^2 + \left(s_{j,k}^{(2)}\right)^2}\right\}}, & \begin{array}{l} j=1,\ldots,m-1, \\ k=1,\ldots,n-1 \end{array} \\[2ex] \dfrac{(1-\alpha) s_{j,k}^{(2)}}{\max\left\{1-\alpha, \left|s_{j,k}^{(2)}\right|\right\}}, & \begin{array}{l} j=m, \\ k=1,\ldots,n-1 \end{array} \end{cases} \quad (35)$$

For the type-II anisotropic TV, $S$ consists of all $\mathbf{q}$ satisfying

$$\begin{aligned} \left|r_{j,k}^{(1)}\right| &\leq \alpha, \quad j=1,\ldots,m-1, k=1,\ldots,n \\ \left|r_{j,k}^{(2)}\right| &\leq \alpha, \quad j=1,\ldots,m, k=1,\ldots,n-1 \\ \left|s_{j,k}^{(1)}\right| &\leq 1-\alpha, \quad j=1,\ldots,m-1, k=1,\ldots,n \\ \left|s_{j,k}^{(2)}\right| &\leq 1-\alpha, \quad j=1,\ldots,m, k=1,\ldots,n-1. \end{aligned} \quad (36)$$

The corresponding projection is given by

$$\begin{aligned} \tilde{r}_{j,k}^{(1)} &= \dfrac{\alpha r_{j,k}^{(1)}}{\max\left\{\alpha, \left|r_{j,k}^{(1)}\right|\right\}}, \quad j=1,\ldots,m-1, k=1,\ldots,n \\ \tilde{r}_{j,k}^{(2)} &= \dfrac{\alpha r_{j,k}^{(2)}}{\max\left\{\alpha, \left|r_{j,k}^{(2)}\right|\right\}}, \quad j=1,\ldots,m, k=1,\ldots,n-1 \\ \tilde{s}_{j,k}^{(1)} &= \dfrac{(1-\alpha) s_{j,k}^{(1)}}{\max\left\{1-\alpha, \left|s_{j,k}^{(1)}\right|\right\}}, \quad j=1,\ldots,m-1, k=1,\ldots,n \\ \tilde{s}_{j,k}^{(2)} &= \dfrac{(1-\alpha) s_{j,k}^{(2)}}{\max\left\{1-\alpha, \left|s_{j,k}^{(2)}\right|\right\}}, \quad j=1,\ldots,m, k=1,\ldots,n-1. \end{aligned} \quad (37)$$

APPENDIX C
GRADIENT CALCULATION

In this Section, before carrying out the derivation, we adopt vectorized notations and regard any matrices or matrix pairs as one-dimensional vectors. This vectorization step is mainly for illustration purposes, and is not necessary in practice. We apologize for this slight abuse of notation, but hopefully this may help keep the derivation concise and thus more accessible to readers. In the following, we use a hat symbol to denote the vectorized matrices. For example, for $\mathbf{x} \in \mathbb{R}^{m \times n} \times \mathbb{R}^{m \times n}$, we have $\hat{\mathbf{x}} \in \mathbb{R}^{2mn \times 1}$. In this case, the linear transmission operator $\mathcal{A}(\cdot)$ can be expressed as a matrix $\mathbf{A}$. We further define some intermediate variables $\hat{\boldsymbol{\rho}} = |\mathbf{A}\hat{\mathbf{x}}| - \sqrt{\hat{\mathbf{y}}}$ and $\hat{\boldsymbol{\xi}} = \mathbf{A}\hat{\mathbf{x}}$. The gradient of $F = 1/2\, \hat{\boldsymbol{\rho}}^T \hat{\boldsymbol{\rho}}$ is calculated according to the chain rule as [48]

$$\frac{\partial F}{\partial \hat{\mathbf{x}}} = \frac{\partial F}{\partial \hat{\boldsymbol{\rho}}} \frac{\partial \hat{\boldsymbol{\rho}}}{\partial \hat{\mathbf{x}}} = \frac{\partial F}{\partial \hat{\boldsymbol{\rho}}} \left( \frac{\partial \hat{\boldsymbol{\rho}}}{\partial \hat{\boldsymbol{\xi}}} \frac{\partial \hat{\boldsymbol{\xi}}}{\partial \hat{\mathbf{x}}} + \frac{\partial \hat{\boldsymbol{\rho}}}{\partial \overline{\hat{\boldsymbol{\xi}}}} \frac{\partial \overline{\hat{\boldsymbol{\xi}}}}{\partial \hat{\mathbf{x}}} \right). \quad (38)$$

The partial derivatives are given by

$$\frac{\partial F}{\partial \hat{\boldsymbol{\rho}}} = \frac{\partial}{\partial \hat{\boldsymbol{\rho}}} \left( \frac{1}{2} \hat{\boldsymbol{\rho}}^T \hat{\boldsymbol{\rho}} \right) = \hat{\boldsymbol{\rho}}^T, \quad (39)$$

$$\begin{aligned} \frac{\partial \hat{\boldsymbol{\rho}}}{\partial \hat{\boldsymbol{\xi}}} &= \frac{\partial}{\partial \hat{\boldsymbol{\xi}}} \left( |\hat{\boldsymbol{\xi}}| - \sqrt{\hat{\mathbf{y}}} \right) \\ &= \frac{\partial \left(|\hat{\boldsymbol{\xi}}|^2\right)^{1/2}}{\partial \left(|\hat{\boldsymbol{\xi}}|^2\right)} \frac{\partial \left(\hat{\boldsymbol{\xi}} \odot \overline{\hat{\boldsymbol{\xi}}}\right)}{\partial \hat{\boldsymbol{\xi}}} \\ &= \frac{1}{2} \mathrm{diag}\left(\frac{1}{|\hat{\boldsymbol{\xi}}|}\right) \mathrm{diag}\left(\overline{\hat{\boldsymbol{\xi}}}\right), \end{aligned} \quad (40)$$

$$\frac{\partial \hat{\boldsymbol{\rho}}}{\partial \overline{\hat{\boldsymbol{\xi}}}} = \overline{\left(\frac{\partial \hat{\boldsymbol{\rho}}}{\partial \hat{\boldsymbol{\xi}}}\right)} = \frac{1}{2} \mathrm{diag}\left(\frac{1}{|\hat{\boldsymbol{\xi}}|}\right) \mathrm{diag}\left(\hat{\boldsymbol{\xi}}\right), \quad (41)$$

$$\frac{\partial \hat{\boldsymbol{\xi}}}{\partial \hat{\mathbf{x}}} = \mathbf{A}, \quad (42)$$

$$\frac{\partial \overline{\hat{\boldsymbol{\xi}}}}{\partial \hat{\mathbf{x}}} = \overline{\mathbf{A}}. \quad (43)$$

The derivation above assumes that $\hat{\boldsymbol{\xi}} = \mathbf{A}\hat{\mathbf{x}}$ has no zero entries, otherwise the gradient is not well-defined. Nevertheless, this problem can be circumvented by assigning a certain complex argument to the zero entries. Substituting (39)-(43) into (38), we have

$$\begin{aligned} \frac{\partial F}{\partial \hat{\mathbf{x}}} &= \frac{1}{2} \hat{\boldsymbol{\rho}}^T \mathrm{diag}\left(\frac{1}{|\mathbf{A}\hat{\mathbf{x}}|}\right) \left( \mathrm{diag}\left(\overline{\mathbf{A}\hat{\mathbf{x}}}\right)\mathbf{A} + \mathrm{diag}\left(\mathbf{A}\hat{\mathbf{x}}\right)\overline{\mathbf{A}} \right) \\ &= \mathcal{R}\left( \hat{\boldsymbol{\rho}}^T \mathrm{diag}\left(\frac{1}{|\mathbf{A}\hat{\mathbf{x}}|}\right) \mathrm{diag}\left(\mathbf{A}\hat{\mathbf{x}}\right) \overline{\mathbf{A}} \right). \end{aligned} \quad (44)$$

Thus, the gradient is

$$\begin{aligned} \nabla F(\hat{\mathbf{x}}) &= \left(\frac{\partial F}{\partial \hat{\mathbf{x}}}\right)^T = \mathcal{R}\left( \mathbf{A}^* \mathrm{diag}\left(\overline{\mathbf{A}\hat{\mathbf{x}}}\right) \mathrm{diag}\left(\frac{1}{|\mathbf{A}\hat{\mathbf{x}}|}\right) \hat{\boldsymbol{\rho}} \right) \\ &= \mathcal{R}\left( \mathbf{A}^* \mathrm{diag}\left(\overline{\mathbf{A}\hat{\mathbf{x}}}\right) \mathrm{diag}\left(\frac{1}{|\mathbf{A}\hat{\mathbf{x}}|}\right) \left(|\mathbf{A}\hat{\mathbf{x}}| - \sqrt{\hat{\mathbf{y}}}\right) \right). \end{aligned} \quad (45)$$

Transforming the above vectorized expression into the original space, we obtain

$$\nabla F(\mathbf{x}) = \mathcal{R}\left\{\mathcal{A}^*\left[\left(\frac{1}{|\mathcal{A}(\mathbf{x})|}\odot\mathcal{A}(\mathbf{x})\right)\odot\left(|\mathcal{A}(\mathbf{x})|-\sqrt{\mathbf{y}}\right)\right]\right\}. \quad (46)$$